\documentclass{iau} 
\usepackage{graphicx}

\title[O dwarf stars in 30\,Doradus] 
{Properties of O dwarf stars in 30\,Doradus}

\author[Carolina Sab\'in-Sanjuli\'an ]   
{Carolina Sab\'in-Sanjuli\'an$^{1,2}$ and the VFTS collaboration   
 }

\affiliation{$^1$ Departamento de F\'isica y Astronom\'ia, Universidad de La Serena, Av. Cisternas 1200 Norte, La Serena, Chile  \\
$^2$ Instituto de Investigaci\'on Multidisciplinar en Ciencia y Tecnolog\'ia, Universidad de La Serena, Ra\'ul Bitr\'an 1305, La Serena, Chile \\ email: {\tt cssj@dfuls.cl} \\[\affilskip]
}

\pubyear{2017}
\volume{329}  
\jname{The lives and death-throes of massive stars}
\editors{A.C. Editor, B.D. Editor \& C.E. Editor, eds.}

\begin{document}

\maketitle

\begin{abstract}
We perform a quantitative spectroscopic analysis of 105 presumably single O dwarf
stars in 30 Doradus, located within the Large Magellanic Cloud. 
We use mid-to-high resolution multi-epoch optical
spectroscopic data obtained within the VLT-FLAMES Tarantula Survey. Stellar and wind parameters are derived by means of the automatic tool \textsc{iacob-gbat}, 
which is based on a large grid of \textsc{fastwind} models. We also benefit from the Bayesian tool \textsc{bonnsai}
to estimate evolutionary masses.
We provide a spectral calibration for the effective temperature of O dwarf stars
in the LMC, deal with the mass discrepancy
problem and investigate the wind properties of the sample.
\end{abstract}

\firstsection 

\section{Introduction}

The VLT-FLAMES Tarantula Survey (VFTS, \cite[Evans et al. 2011]{evans11}) is an ESO large programme that has obtained mid-to-high resolution multi-epoch optical 
spectra of hundreds of O-type stars in 30\,Doradus, the largest H\,\textsc{ii} region in the Local Group, located in the Large Magellanic Cloud. VFTS was developed aiming at the study of 
rotation, 
binarity and wind properties of an unprecedented number of massive stars within the same star-forming region. Within this project, a huge effort has been made in the study of O-type stars, 
including the 
analysis of multiplicity (\cite[Sana et al. 2013]{sana13}), spectral classifications (\cite[Walborn et al. 2014]{w14}) and the distribution of rotational velocities 
(\cite[Ram\'irez-Agudelo et al. 2013, 2015]{oscar13,oscar15}). These studies are essential for the estimation of stellar parameters and chemical abundances of the complete O-type sample, which will 
allow to investigate fundamental questions in stellar and cluster evolution. To date, the quantitative study of the O-type sample is about to be complete (see 
\cite[Sab\'in-Sanjuli\'an et al. 2014, 2017]{cssj14,cssj17}; \cite[Ram\'irez-Agudelo et al. 2017]{oscar17}), as well as the determination of nitrogen abundances (\cite[Grin et al. 2016]{grin16}, 
Sim\'on-D\'iaz et al. in prep.)

In this work, we study the properties of a sample of O stars close to the ZAMS, those with luminosity classes IV and V.

\section{Observations and data sample}

We have used spectroscopic data obtained by means of the Medusa mode of the FLAMES spectrograph at the VLT (Paranal, Chile). In addition, $I$-band images obtained with the $HST$ (GO12499, P.~I.: 
D.~J.~Lennon) were 
utilized to evaluate possible contamination in the Medusa fibres.

We have selected a sample of 105 likely single and unevolved O stars based on the multiplicity properties derived by \cite[Sana et al. (2013)]{sana13} and the spectral classifications 
by \cite[Walborn et al. (2014)]{w14}. Our sample includes O stars with luminosity classes V and IV, as well as Vz, V-III (uncertain classification) and III-IV (interpolation between 
III and IV).

\section{Quantitative spectroscopic analysis} \label{sec:qsa}

The quantitative analysis of the spectroscopic data for our sample of O dwarfs was performed by means of the IACOB Grid-Based Automatic Tool (\textsc{iacob-gbat}, see 
\cite[Sim\'on-D\'iaz et al. 2011]{ssd11}, \cite[Sab\'in-Sanjuli\'an et al. 2014]{cssj14}). \textsc{iacob-gbat} is based on a grid of $\sim$190\,000 precomputed \textsc{fastwind} stellar atmosphere 
models (\cite[Santolaya-Rey 
et al. 1997]{sr97}, 
\cite[Puls et al. 2005]{puls05}) and a $\chi^2$ algorithm that compares observed and synthetic H and He line profiles. 

Absolute magnitudes in the $V$-band calculated by \cite[Ma\'iz-Apell\'aniz et al. (2014)]{jma14} and projected rotational velocities by \cite[Ram\'irez-Agudelo et al. (2013)]{oscar13} were used. 

We estimated mean values and uncertainties for effective temperature T$_{\rm eff}$, surface gravity $g$, helium abundance $Y(He)$, stellar radius $R$, luminosity $L$, spectroscopic mass $M_{\rm sp}$ 
and wind-strength $Q$-parameter (defined as $Q$\,=\,$\dot{M}(R\,v_{\infty})^{-3/2}$, where 
$\dot{M}$ is the mass-loss rate and $v_{\infty}$ the wind terminal velocity). Evolutionary masses ($M_{\rm ev}$) were obtained by means of the Bayesian tool \textsc{bonnsai} (\cite[Schneider et al. 
2014]{bonnsai}), 
which used
evolutionary models by \cite[Brott et al. (2011)]{brott11} and \cite[K\"ohler et al. (2015)]{kohler15}. 

Most of the stars in our sample showed strong nebular contamination, but only 11 critical cases had to be analyzed using nitrogen lines (HHeN). This diagnostic was also used for 9 cases 
with too weak or 
inexistent He\,\textsc{i} lines. We could only provide upper limits for log\,$Q$ for about 70\% of the sample due to thin winds. In addition, possible/confirmed contamination in fibre was present in 
several stars, which could have altered the determination of physical parameters. A few cases were found with too low helium abundances and too high gravities, a possible indication of undetected 
binarity. 
\section{Spectral calibration} 

\begin{figure}[h!!!]
\begin{center}
 \includegraphics[scale=0.5,angle=90]{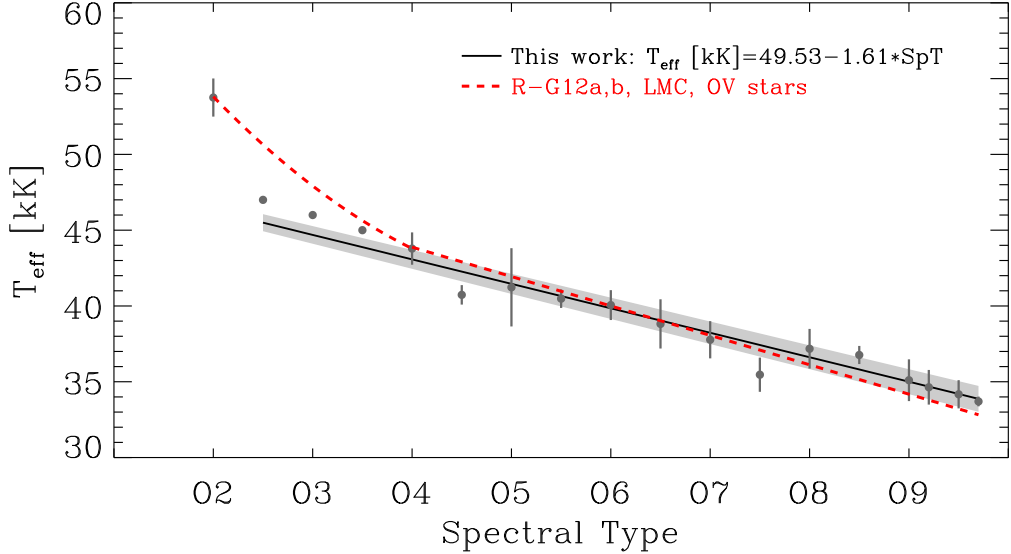} 
 \caption{Effective temperature calibration compared to that derived by \cite[Rivero-Gonz\'alez et al. (2012a,b)]{rivero12a,rivero12b}.  For each spectral type, the dark grey circles and bars 
represent the
mean value and standard deviation of our T$_{\rm eff}$ estimates. Our linear fit is represented by the black line, and the
grey zones indicate the associated uncertainties. O2 stars have been excluded from the fit due to their possible contamination in fibre and/or undetected binarity. }
   \label{fig:calibration}
\end{center}
\end{figure}

Figure~\ref{fig:calibration} represents our T$_{\rm eff}$ scale as a function of spectral types. We perform a linear fit to the points excluding O2 stars, since they show indications of 
undetected binarity and/or contamination in fibre. When comparing with the most recent and complete spectral calibration for O dwarfs in the LMC before this work (\cite[Rivero-Gonz\'alez et al. 
2012a,b]{rivero12a,rivero12b}, who used HHeN diagnostics to analyse optical spectra of 25 stars, including 16 dwarfs), we found that there is an excellent agreement between both scales for spectral 
types later than O4. In the earliest regime, Rivero-Gonz\'alez et al. utilized a quadratic fit obtaining hotter temperatures than ours, although the estimates for our O2 stars agree with their 
calibration. However, we cannot reach a firm conclusion about the necessity of changing the slope in the O2-O4 regime due to the small number of observed stars and their very likely 
binarity/multiplicity. A larger sample of O-type stars earlier than O4, as well as information on their possible composite nature are necessary.

\section{Mass discrepancy }

The mass discrepancy problem consists on a lack of agreement between spectroscopic and evolutionary masses. Since its identification by \cite[Herrero et al. 
(1992)]{h92}, several explanations have been proposed but it remains unresolved due to the diversity of results in different environments.

\begin{figure}[h!!!]
\begin{center}
 \includegraphics[scale=0.5,angle=90,trim=0mm 0mm 0mm 140mm,clip]{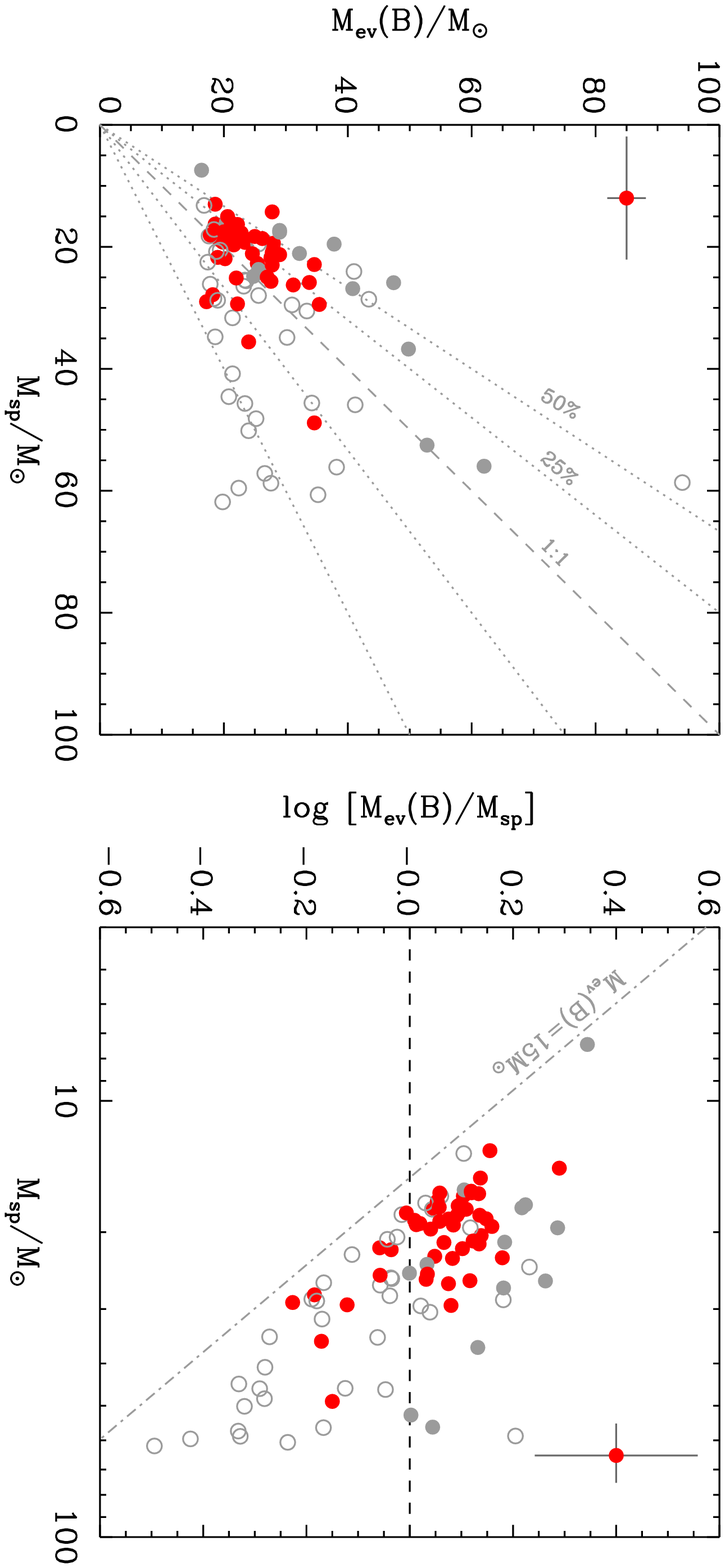} 
 \caption{Logarithmic ratio of evolutionary masses from
  \textsc{bonnsai} ($M_{\rm ev}(B)$) to spectroscopic masses ($M_{\rm
    sp}$). Grey filled symbols correspond to stars analyzed with nitrogen lines and no indications of binarity or contamination in fibre. Open symbols 
are those stars with 
possible undetected binarity or contamination in fibre. Typical error bars are given in the upper-right corner. The dashed
  line indicates points where $M_{\rm ev}(\rm B)$\,=\,15\,M$_{\odot}$. O2 stars are excluded.}
   \label{fig:mass_discrep}
\end{center}
\end{figure}

We compare our derived spectroscopic  and evolutionary  masses in Figure~\ref{fig:mass_discrep}. We find a certain trend in the distribution: for masses 
below 20~M$_{\odot}$, we can only find stars with $M_{\rm ev}$\,$>$\,$M_{\rm sp}$, i.e., a positive discrepancy, while for higher masses both positive and negative discrepancies are present.
Nevertheless, this trend could be explained by a selection bias in our sample. All the O dwarfs in this work have masses systematically above 15~M$_{\odot}$ (see indication in 
Fig.~\ref{fig:mass_discrep}). As a consequence, all stars with a negative mass discrepancy and masses below 15~M$_{\odot}$ are not present. To investigate this effect and reach a clearer conclusion 
about the mass discrepancy in the VFTS sample, results from the ongoing quantitative analysis of the early B-dwarfs should be included.

\section{Wind properties}

\begin{figure}[h!!!]
\begin{center}
 \includegraphics[scale=0.5,angle=90,trim=0mm 0mm 0mm 0mm,clip]{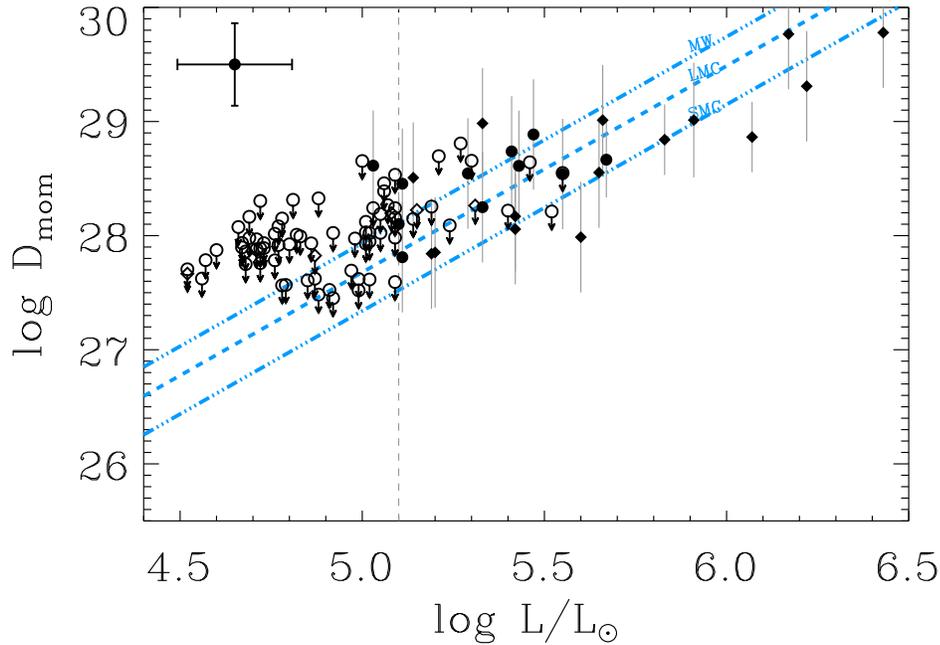} 
 \caption{Wind-momentum\,--\,Luminosity Relationship (WLR) for our sample compared with theoretical predictions from \cite[Vink et al. (2001)]{vink01} for the metallicities of the Milky Way and the 
Magellanic Clouds. Open symbols represent stars with upper limits on log\,$Q$ (hence $D_{\rm mom}$). Diamonds correspond to stars analyzed using nitrogen lines. 
Typical uncertainties are shown in the
upper-left corner, and the vertical dashed line indicates log\,$L/L_{\odot}$\,=\,5.1.}
   \label{fig:wind}
\end{center}
\end{figure}

In this study, we estimated the logarithm of the wind-strength $Q$-parameter (see Sect.~\ref{sec:qsa}) of the sample stars. If terminal velocities were available, log\,$Q$ could be used to estimate 
mass-loss rates. Nevertheless, we lack UV spectroscopic data to derive directly terminal velocities. For this reason, we have followed the approach used by \cite[Mokiem et al. 
(2007a,b)]{mokiem07a,mokiem07b} to estimate the escape velocity $v_{esc} = 2gR(1-\Gamma)^{1/2}$ and the terminal velocity $v_{\infty}/v_{esc} = 2.65\,Z^{0.13}$, where  $\Gamma$ is the 
Eddington factor and $Z$ the metallicity corresponding to the LMC.  The wind momentum was calculated via $D_{\rm mom}$\,=\,$\dot{M}v_{\infty}R^{1/2}$, and is represented 
in Figure~\ref{fig:wind} as a function of stellar luminosity. 
Two different behaviours are present in the distribution: 
\begin{itemize}
 \item At luminosities higher than log\,$L/L_{\odot}$\,$\sim$\,5.1, winds were constrained for most of the stars. Most of them show consistency with the linear trend predicted by \cite[Vink et 
al. (2001)]{vink01} within the error bars. However, a large dispersion is found, showing some stars above the theoretical WLR for Galactic metallicity and some others below the prediction for the 
SMC. The 
first case could be explained by unclumped wind models and large uncertainties on the estimated terminal velocities, but no suitable explanation can be found for the second one.
\item At luminosities lower than log\,$L/L_{\odot}$\,$\sim$\,5.1, only upper limits could be given for log\,$Q$ and therefore $D_{\rm mom}$. To constrain such thin 
winds, UV spectroscopy is necessary.

\end{itemize}

%
%
%
%
%
%
%


\begin{thebibliography}{}

\bibitem[\protect\citeauthoryear{Brott et al.}{2011}]{brott11} Brott I., et al., 2011, A\&A, 530, A115 
\bibitem[\protect\citeauthoryear{Evans et al.}{2011}]{evans11} Evans C.~J., et al., 2011, A\&A, 530, A108 
\bibitem[\protect\citeauthoryear{Grin et al.}{2016}]{grin16} Grin N.~J., et al., 2016, arXiv, arXiv:1609.00197 
\bibitem[\protect\citeauthoryear{Herrero et al.}{1992}]{h92} Herrero A., Kudritzki R.~P., Vilchez J.~M., Kunze D., Butler K., Haser S., 1992, A\&A, 261, 209 
\bibitem[\protect\citeauthoryear{K{\"o}hler et al.}{2015}]{kohler15} K{\"o}hler K., et al., 2015, A\&A, 573, A71 
\bibitem[\protect\citeauthoryear{Ma{\'{\i}}z Apell{\'a}niz et al.}{2014}]{jma14} Ma{\'{\i}}z Apell{\'a}niz J., et al., 2014, A\&A, 564, A63 
\bibitem[\protect\citeauthoryear{Mokiem et al.}{2007}]{mokiem07a} Mokiem M.~R., et al., 2007, A\&A, 465, 1003 
\bibitem[\protect\citeauthoryear{Mokiem et al.}{2007}]{mokiem07b} Mokiem M.~R., et al., 2007, A\&A, 473, 603 
\bibitem[\protect\citeauthoryear{Puls et al.}{2005}]{puls05} Puls J., Urbaneja M.~A., Venero R., Repolust T., Springmann U., Jokuthy A., Mokiem M.~R., 2005, A\&A, 435, 669 
\bibitem[\protect\citeauthoryear{Ram{\'{\i}}rez-Agudelo et al.}{2013}]{oscar13} Ram{\'{\i}}rez-Agudelo O.~H., et al., 2013, A\&A, 560, A29
\bibitem[\protect\citeauthoryear{Ram{\'{\i}}rez-Agudelo et al.}{2015}]{oscar15} Ram{\'{\i}}rez-Agudelo O.~H., et al., 2015, A\&A, 580, A92
\bibitem[\protect\citeauthoryear{Ram{\'{\i}}rez-Agudelo et al.}{2017}]{oscar17} Ram{\'{\i}}rez-Agudelo O.~H., et al., 2017, arXiv, arXiv:1701.04758 
\bibitem[\protect\citeauthoryear{Rivero Gonz{\'a}lez et al.}{2012}]{rivero12a} Rivero Gonz{\'a}lez J.~G., Puls J., Najarro F., Brott I., 2012, A\&A, 537, A79 
\bibitem[\protect\citeauthoryear{Rivero Gonz{\'a}lez et al.}{2012}]{rivero12b} Rivero Gonz{\'a}lez J.~G., Puls J., Massey P., Najarro F., 2012, A\&A, 543, A95 
\bibitem[\protect\citeauthoryear{Sab{\'{\i}}n-Sanjuli{\'a}n et al.}{2014}]{cssj14} Sab{\'{\i}}n-Sanjuli{\'a}n C., et al., 2014, A\&A, 564, A39 
\bibitem[\protect\citeauthoryear{Sab{\'{\i}}n-Sanjuli{\'a}n et al.}{2017}]{cssj17} Sab{\'{\i}}n-Sanjuli{\'a}n C., et al., 2017, A\&A, 601, A79 
\bibitem[\protect\citeauthoryear{Sana et al.}{2013}]{sana13} Sana H., et al., 2013, A\&A, 550, A107 
\bibitem[\protect\citeauthoryear{Santolaya-Rey, Puls, \& Herrero}{1997}]{sr97} Santolaya-Rey A.~E., Puls J., Herrero A., 1997, A\&A, 323, 488 
\bibitem[\protect\citeauthoryear{Schneider et al.}{2014}]{bonnsai} Schneider F.~R.~N., Langer N., de Koter A., Brott I., Izzard R.~G., Lau H.~H.~B., 2014, A\&A, 570, A66 
\bibitem[\protect\citeauthoryear{Sim{\'o}n-D{\'{\i}}az et al.}{2011}]{ssd11} Sim{\'o}n-D{\'{\i}}az S., Castro N., Herrero A., Puls J., Garcia M., Sab{\'{\i}}n-Sanjuli{\'a}n C., 2011, JPhCS, 328, 
012021 
\bibitem[\protect\citeauthoryear{Vink, de Koter, \& Lamers}{2001}]{vink01} Vink J.~S., de Koter A., Lamers H.~J.~G.~L.~M., 2001, A\&A, 369, 574 
\bibitem[\protect\citeauthoryear{Walborn et al.}{2014}]{w14} Walborn N.~R., et al., 2014, A\&A, 564, A40

\end{thebibliography}
\end{document}